\documentstyle[12pt]{article}

\title{
\begin{flushright}
{\normalsize Yaroslavl State University\\
             Preprint YARU-HE-95/01\\
             hep-ph/9503239} \\[5mm]
\end{flushright}
The Minimal Quark-Lepton Symmetry Model and the Limit
       on Z'-mass.}

\author{A.D.~Smirnov\thanks{E-mail: phystheo@univ.uniyar.ac.ru}\\
{\small\it Division of Theoretical Physics, Department of Physics,}\\
{\small\it Yaroslavl State University, Sovietskaya 14,}\\
{\small\it 150000 Yaroslavl, Russia.}}

\date{}

\begin{document}

\maketitle

\begin{abstract}

A minimal extension of the Standard Model containing the
four-color quark-lepton symmetry is proposed and discussed.
The existence of a rather light extra $Z'$-boson originated
from the four-color quark-lepton symmetry is shown to be
compatible with the current electroweak data. The cross
sections $\sigma(e^+ e^- \rightarrow \gamma, Z, Z' \rightarrow
\bar f f)$ are calculated, and their deviations from the SM
predictions are shown to be significant at $\sqrt s \ge 200 \, GeV$
and available for the measurements at the LEP200 and future
colliders.

\end{abstract}

\vglue 15mm

\begin{center}
{\it To be published in Phys. Lett. B}
\end{center}

\newpage

The successful unification of the electromagnetic and weak interactions
by the Standard Model (SM) and the idea of the possible existence of
the more large symmetries at high energies stimulate the search for
a new physics at the present and future colliders. It seems that
high symmetries will manifest themselves consecutively one after
an other as the energies of colliders increase.
The $SU(2) \times U(1)$-symmetry of the SM is, probably, only the
first such symmetry, what is the next one then?
To answer this question
it seems reasonable to investigate various minimal extensions of the SM
by adding to it some additional symmetries such as the right
$SU_R(2)$-symmetry in the $SU_L(2) \times SU_R(2) \times U(1)$-models,
the supersymmetry in the Minimal Supersymmetric Standard Model
(MSSM), etc. One of such symmetries possibly existing
in nature and being worthy of the detailed investigation now is the
four-color quark-lepton symmetry regarding the lepton number as the
fourth color \cite{PSm}.

In this work we propose the minimal quark-lepton symmetry model
of the unification of the strong and electroweak interactions (MQLS-model)
which is the minimal extension of the SM containing the four-color
quark-lepton symmetry. We discuss some features of the extra Z'-bosons
originated from this symmetry.

The model to be discussed here is based on the $SU_V(4) \times SU_L(2)
\times U_R(1)$-group as the minimal group containing the four-color
symmetry of quarks and leptons. In this model the quarks $Q_{p a \alpha}
= \psi_{p a \alpha}$,
$a$ = 1, 2, $\alpha$ = 1, 2, 3 and the corresponding leptons $\ell_{p a}
= \psi_{p a 4}$ in each generation of the number $p$ = 1, 2, 3, $\ldots$
form the four-color
fundamental quartet $\psi_{p a A}$, $A$ = 1, 2, 3, 4 of the $SU_V(4)$-group.
Under the $SU_L(2) \times U_R(1)$-group
the left fermions are the doublets with $Y_L$ = 0 and the
right fermions are the singlets with $Y_R$ = $\pm$ 1 for the ``up''
($a$ = 1) and ``down'' ($a$ = 2) fermions respectively.
For three generations the basic up- and down- fermion $SU_V(4)$-quartets are

\begin{displaymath}
\psi'_{p 1 A} \; : \quad
\left ( \begin{array}{c} u'_\alpha \\ \nu'_e \end{array} \right ) , \;
\left ( \begin{array}{c} c'_\alpha \\ \nu'_\mu \end{array} \right ) , \;
\left ( \begin{array}{c} t'_\alpha \\ \nu'_\tau \end{array} \right ) , \;
\cdots
\end{displaymath}

\begin{displaymath}
\psi'_{p 2 A} \; : \quad
\left ( \begin{array}{c} d'_\alpha \\ {e^-}' \end{array} \right ) , \;
\left ( \begin{array}{c} s'_\alpha \\ {\mu^-}' \end{array} \right ) , \;
\left ( \begin{array}{c} b'_\alpha \\ {\tau^-}' \end{array} \right ) , \;
\cdots
\end{displaymath}

\noindent where the basic quark and lepton fields
${Q'}^{L,R}_{p a \alpha}$,
${\ell'}^{L,R}_{p a}$
can be written, in general, as superpositions

\begin{eqnarray}
{Q'}^{L,R}_{p a \alpha} = \sum_q \left ( A^{L,R}_{Q_a} \right )_{p q}
Q^{L,R}_{q a \alpha} , \qquad
{\ell'}^{L,R}_{p a} = \sum_q \left ( A^{L,R}_{\ell_a} \right )_{p q}
\ell^{L,R}_{q a}  \nonumber
\end{eqnarray}

\noindent of mass eigenstates
$Q^{L,R}_{q a \alpha}$, $\ell^{L,R}_{q a}$. Here
$A^{L,R}_{Q_a}$ and $A^{L,R}_{\ell_a}$ are unitary matrices
diagonalizing the mass matrices of quarks and leptons respectively.
The electric charges of quarks and leptons are related to the
generators of the group by

\begin{displaymath}
Q_{L,R}^{em} \, = \, \sqrt{\frac{2}{3}} \, t_{15}^{L,R} \, + \,
\frac{\tau_3^L}{2} \, + \, \frac{Y^R}{2} \, ,
\end{displaymath}

\noindent where $t_{15}$, $\tau_3/2$ are the corresponding generators,
$\tau_3$ is the Pauli matrix.

According to the structure of the group the gauge sector consists
of 19 fields $A_\mu^i$, $i = 1, 2, \ldots, 15$, $W_\mu^k$,
$k = 1, 2, 3$ and $B_\mu$. The first eight of them are the gluons
$G_\mu^j$ = $A_\mu^j$, $j = 1, 2, \ldots, 8$, the next six fields
form the triplets of the leptoquarks $V_{\alpha \mu}^\pm$, $\alpha = 1, 2, 3$
with the electric charges $Q_V^{em} = \pm 2/3$, $W^1_\mu$, $W^2_\mu$
form the $W^\pm$-bosons in a usual way and the remained fields
$A^{15}_\mu$, $W^3_\mu$, $B_\mu$ form the photon, the Z-boson
and an extra Z'-boson.

The electromagnetic field $A_\mu$ is related to $A^{15}_\mu$, $W^3_\mu$,
$B_\mu$ by

\begin{displaymath}
A_\mu = s_S A^{15}_\mu + \sqrt{1 - s_W^2 - s_S^2} B_\mu + s_W W_\mu^3 ,
\end{displaymath}

\noindent and two orthogonal to $A_\mu$ fields $Z_{1 \mu}$ and $Z_{2 \mu}$
can be written as

\begin{eqnarray}
Z_{1 \mu} & = & - t_W \big (s_S A^{15}_\mu + \sqrt{1 - s_W^2 - s_S^2}
B_\mu \big ) + c_W W_\mu^3 , \nonumber \\
Z_{2 \mu} & = & \big (\sqrt{1 - s_W^2 - s_S^2} A^{15}_\mu -
s_S B_\mu \big )/c_W , \nonumber
\end{eqnarray}

\noindent where $s_{W,S} = \sin \theta_{W,S}$, $c_W = \cos \theta_W$,
$t_W = \tan \theta_W$. The angles $\theta_W$ and $\theta_S$ of the weak
and strong mixings are defined as

\begin{eqnarray}
s_W^2 & = & \frac{\alpha(m)}{\alpha_W(m)} , \label{eq:sW} \\
s_S^2 & = & \frac{2}{3} \, \frac{\alpha(m)}{\alpha_{15}(m)} \, = \,
\frac{2}{3} \, \frac{\alpha(m)}{\alpha_S(m)} \, \left [ 1 +
\frac{\alpha_S(m)}{2 \pi} \, \left ( b \, \ln \frac{M_C}{m} +
b_{15} \, \ln \frac{M'}{m} \right ) \right ],
\label{eq:sS}
\end{eqnarray}

\noindent where $\alpha(m)$, $\alpha_W(m)$, $\alpha_S(m)$ are the
electromagnetic, weak and strong coupling constants at the scale $m$,
$M_C$ is the mass scale of the $SU_V(4)$-symmetry breaking,
$M'$ is the possible intermediate mass scale of $U_{15}(1)$-symmetry
breaking, $b = b_S - b_{15}$, $b_S$ and $b_{15}$ are the group constants
for the group $SU_c(3)$ and $U_{15}(1)$ respectively. Taking into
account the gauge and fermion fields gives $b_S = 11 - (4/3) n_G$,
$b_{15} = - (4/3) n_G$ and $b = 11$, $n_G$ is the number of fermion
generations with masses below $M_C$.
The last equality in~(\ref{eq:sS}) is obtained by the elimination of
the $SU_V(4)$ unified gauge coupling constant
$\alpha_4(M_C) = g_4^2 / 4 \pi$
from the one-loop approximation relations

\begin{equation}
\alpha_{S,15}(m) \, = \, \alpha_4(M_C)/\big ( 1 - \frac{\alpha_4(M_C)}
{2 \pi} \, b_{S,15} \, \ln \frac{M_C}{m} \big ) \label{eq:aS15}
\end{equation}

\noindent between $\alpha_4(M_C)$
and the $A^{15}$-interaction constant
$\alpha_{15}(m)$ and $\alpha_S(m)$.

The interaction of the gauge fields with the fermions has the form

\begin{eqnarray}
{\cal L}_\psi^{gauge} & = & \frac{g_4}{\sqrt 2} \big\lbrace V^\alpha_\mu
\big [ \bar Q^L_{p a \alpha} \gamma^\mu \big ( K^L_a \big )_{p q}
\ell^L_{q a} +
\bar Q^R_{p a \alpha} \gamma^\mu \big ( K^R_a \big )_{p q}
\ell^R_{q a} \big ] + h.c. \big\rbrace \nonumber \\
& + & \frac{g_2}{\sqrt 2} \big\lbrace  W^+_\mu \big [
\bar Q^L_{p 1 \alpha} \gamma^\mu \big ( C_Q \big )_{p q} Q^L_{q 2 \alpha} +
\bar \ell^L_{p 1} \gamma^\mu \big ( C_\ell \big )_{p q} \ell^L_{q 2}
\big ] + h.c. \big\rbrace  \nonumber \\
& + & g_{st} G_\mu^j \big ( \bar Q \gamma^\mu t_j Q \big ) -
|e| A_\mu \big ( \bar \psi \gamma^\mu Q^{em} \psi \big ) +
{\cal L}_{N C}^{gauge}.
\label{eq:Lint}
\end{eqnarray}

\noindent Here the first term describes the interaction
of leptoquarks with quarks and leptons by the constant $g_4$ related to
$\alpha_S(m)$ by~(\ref{eq:aS15}). This interaction contains, in general,
the new generation mixing due to the matrices $K^{L,R}_a =
(A^{L,R}_{Q_a})^+ A^{L,R}_{\ell_a}$.
These matrices should be extracted from the experiments with the
leptoquarks. Some restrictions on these matrices and on the leptoquark
masses resulting from the rare $K$, $\pi$ and $B$ decays in the case of
$K^L_a = K^R_a$ have been
investigated recently in Ref.~\cite{KM,VW}.
The second term in~(\ref{eq:Lint}) describes
the weak charged current interaction of $W^\pm$-bosons with quarks or
leptons by the constant $g_2$ related to the Fermi constant $G_F$
and $m_W$ in a usual way.
This interaction contains the well known Cabibbo-Kobayashi-Maskawa
mixing of the quarks due to the CKM matrix
$C_Q = (A^L_{Q_1})^+ A^L_{Q_2}$
and, in general, the analogous mixing in the lepton sector due to the
lepton mixing matrix
$C_\ell = (A^L_{\ell_1})^+ A^L_{\ell_2}$.
The next two terms are the
QCD- and QED- interactions. The neutral current interaction with the
gauge fields ${\cal L}_{N C}^{gauge}$ can be written as

\begin{equation}
{\cal L}_{N C}^{gauge} = - Z_\mu J^Z_{\mu} - Z'_\mu J^{Z'}_{\mu} ,
\label{eq:Lnc}
\end{equation}

\noindent where

\begin{eqnarray}
Z_\mu & = & Z_{1 \mu} \cos \theta_m + Z_{2 \mu} \sin \theta_m , \nonumber \\
{Z'}_\mu & = & - Z_{1 \mu} \sin \theta_m + Z_{2 \mu} \cos \theta_m  \nonumber
\end{eqnarray}

\noindent are the mass eigenstate fields and

\begin{eqnarray}
J_\mu^Z & = & J_\mu^{Z_1} \cos \theta_m + J_\mu^{Z_2} \sin \theta_m ,
\label{eq:JZ} \\
J_\mu^{Z'} & = & - J_\mu^{Z_1} \sin \theta_m + J_\mu^{Z_2} \cos \theta_m ,
\label{eq:JZ'} \\
J^{Z_1}_{\mu} & = & \frac{|e|}{s_W c_W} \, \big ( J^{3 L}_{\mu} -
   s_W^2 J^{em}_\mu \big ), \label{eq:JZ1} \\
J^{Z_2}_\mu & = & \frac{|e|}{s_S c_W \sqrt{1 - s_W^2 - s_S^2}} \,
\Big [ c_W^2 \sqrt{\frac{2}{3}} J^{15}_{\mu} - s_S^2 \big (
J^{em}_\mu - J^{3 L}_\mu \big ) \Big ]
\label{eq:JZ2}
\end{eqnarray}

\noindent with the currents
$J^{em}_\mu = (\bar \psi_{p a A} \gamma_\mu Q_{a A}^{em} \psi_{p a A})$,
$J^{3 L}_\mu = \frac{1}{2} (\bar \psi_{p a A} \gamma_\mu (1 + \gamma_5)
(\tau_3/2)_{a a} \psi_{p a A})$,
$J^{15}_\mu = (\bar \psi_{p a A} \gamma_\mu (t_{15})_{A A} \psi_{p a A})$.
The $Z_1$-current (\ref{eq:JZ1}) is the usual neutral current of
the Standard Model, but the structure of the $Z_2$-current (\ref{eq:JZ2})
is specified by the model under consideration. The $Z-Z'$-mixing
angle $\theta_m$ is defined by the symmetry breaking mechanism of
the model and is found to be small.

The Higgs sector of the model is taken in the simplest way and consists
of the four multiplets (4, 1, 1), (1, 2, 1), (15, 2, 1), (15, 1, 0) of
$SU_V(4) \times SU_L(2) \times U_R(1)$-group with the vacuum expectation
values (VEV's) $\langle \phi_A^{(1)} \rangle = \delta_{A 4} \eta_1 /
\sqrt 2$, $\langle \phi_a^{(2)} \rangle = \delta_{a 2} \eta_2 / \sqrt 2$,
$\langle \phi_{i a}^{(3)} \rangle = \delta_{i 15} \delta_{a 2} \eta_3$,
$i$ = 1, 2, $\ldots$, 15
and $\langle \phi_i^{(4)} \rangle = \delta_{i 15} \eta_4$ respectively.
The field $\phi^{(1)}$ breaks the $SU_V(4)$- symmetry down to $SU_C(3)$
giving the masses to the $Z'$- boson and to the leptoquarks.
The $SU_L(2)$- doublet $\phi^{(2)}$ breaks the $SU_L(2)$- symmetry
in the usual way and gives the equal masses to the fermions belonging
to the same generation. The main function of the $\phi^{(3)}$ multiplet
is to split the masses of the quarks and leptons in each generation.
The $\phi^{(4)}$ multiplet breaks the $SU_V(4)$- symmetry down to
$SU_C(3) \times U_{15}(1)$ contributing to the leptoquark masses and
splitting them from the $Z'$- mass.
After breaking the symmetry in such a way the masses of quarks
and leptons are defined
by VEV's $\eta_2$, $\eta_3$ and by Yukawa coupling constants and can be
arbitrary just as they are in the Standard Model, the photon and the gluons
are still massless but all the other gauge fields acquire the masses.

The model has, in general, three mass scales determined by the VEV's
$\eta = \sqrt{\eta_2^2 + \eta_3^2}$, $\eta_1$ and $\eta_4$ respectively.
The first mass scale is the usual mass scale of the Standard Model
$\eta = (\sqrt 2 G_F)^{-1/2} \simeq 250 \, GeV$, the second one is
the possible intermediate mass scale $M' \sim m_{Z'}$ of
$U_{15}(1)$-symmetry breaking
and the third one is the mass scale $M_C \sim m_V$
of $SU_V(4)$- symmetry breaking. The low limit on $M_C$ can be
about a few hundreds $TeV$ or, possibly, somewhat lower~\cite{KM, VW}
in dependence on the character of the $K^{L,R}_a$ mixing in the
leptoquark interaction in~(\ref{eq:Lint}).

In the case of $\eta_4 \gg \eta_1$ the symmetry breaking has the
three stage form

\begin{eqnarray}
SU_V(4) \times SU_L(2) \times U_R(1)
\stackrel{\eta_4}{\longrightarrow}
SU_C(3) \times U_{15}(1) \times SU_L(2) \times U_R(1) \nonumber \\
\stackrel{\eta_1}{\longrightarrow}
SU_C(3) \times SU_L(2) \times U(1)
\stackrel{\eta}{\longrightarrow}
SU_C(3) \times U_{em}(1).
\nonumber
\end{eqnarray}

\noindent Because the VEV $\eta_4$ contributes only to the masses
of the leptoquarks the $Z'$- boson can be rather light
($m_{Z'} \sim M' \sim \eta_1$) in this case with the leptoquarks being
sufficiently heavy ($m_V \sim M_C \sim \eta_4$).

In another limiting case of $\eta_4 = 0$ the model has only two mass
scales $\eta$ and $M' \sim M_C \sim \eta_1$,
the symmetry breaking has two stages

\begin{displaymath}
SU_V(4) \times SU_L(2) \times U_R(1)
\stackrel{\eta_1}{\longrightarrow}
SU_C(3) \times SU_L(2) \times U(1)
\stackrel{\eta}{\longrightarrow}
SU_C(3) \times U_{em}(1)
\end{displaymath}

\noindent and $Z'$-boson must be as heavy as the leptoquarks are:
$m_{Z'} \sim m_V \sim M_C \sim M' \sim \eta_1$.

Irrespective of the hierarchy of VEV's $\eta_1$ and $\eta_4$ the
model predicts the relation between the masses of the $W$-, $Z$- and
$Z'$- bosons

\begin{equation}
\big ( \mu^2 - \rho_0 \big ) \big ( \rho_0 - 1 \big ) = \rho^2_0 \sigma^2 ,
\label{eq:mr}
\end{equation}

\noindent where $\mu \equiv m_{Z'} / m_Z$, $\rho_0 \equiv m_W^2 / m_Z^2
c_W^2$ and

\begin{equation}
\sigma = \frac{s_W s_S}{\sqrt{1 - s_W^2 - s_S^2}}. \label{eq:s}
\end{equation}

\noindent Simultaneously the model gives
for the $Z$ - $Z'$ mixing angle $\theta_m$
the expression

\begin{equation}
\sin \theta_m = \Big [ 1 + \Big ( \frac{\rho_0 \sigma}{\rho_0 - 1} \Big )^2
\Big ]^{-1/2}  \label{eq:sint}
\end{equation}

\noindent For $\theta_m \ll 1$ and $\rho_0 \simeq 1$
we also obtain from (\ref{eq:mr}), (\ref{eq:sint}) that $\theta_m
\simeq \sigma \, m_Z^2 / m_{Z'}^2$.

It should be noted for comparision that the extended gauge model based
on the $SU_L(2) \times U(1) \times U'(1)$- group contains the $Z'$- boson
mass and the $Z$ - $Z'$ mixing angle as the independent and arbitrary
parameters because the corresponding to the additional group $U'(1)$
coupling constant is arbitrary in this model. Unlike this the $U_{15}(1)$
coupling constant $g_{15}$ of the MQLS model is related to the strong
coupling constant $\alpha_S$ by~(\ref{eq:aS15}), which leads as a result
to the relations~(\ref{eq:mr}),~(\ref{eq:s}).

The Yukawa interaction of the fermions with the scalar fields contains
only the $SU_L(2)$- doublets $\phi^{(2)}$ and $\phi^{(3)}_i$ and has,
in general, the form

\begin{equation}
{\cal L}_\psi^{Yukawa} = - \bar \psi'^L_{p a A} \big [
\big ( h_b \big )_{p q} \phi^{(2) b}_a \delta_{A B} +
\big ( h'_b \big )_{p q} \phi^{(3) b}_{i a} \big ( t_i \big )_{A B}
\big ] \psi'^R_{q b B} + h.c. ,
\label{eq:LYukawa}
\end{equation}

\noindent where
$\phi^{(2) 2}_a = \phi^{(2)}_a$,
$\phi^{(2) 1}_a = \varepsilon_{a c} (\phi^{(2)}_c)^*$,
$\phi^{(3) 2}_{i a} = \phi^{(3)}_{i a}$,
$\phi^{(3) 1}_{i a} = \varepsilon_{a c} (\phi^{(3)}_{i c})^*$,
$i$ = 1, 2, $\ldots$, 15, $\varepsilon_{a c}$ is antisymmetrical symbol,
$h_b$ and $h'_b$ are arbitrary matrices.

After breaking the symmetry the
Lagrangian~(\ref{eq:LYukawa}) gives the arbitrary masses to the quarks and
leptons splitting them in each generation. The remained part of the
Lagrangian describes the interaction of the fermions with the scalar
fields enterring into $\phi^{(2)}$- and $\phi^{(3)}$- multiplets.
Among these fields there are the $SU_L(2)$-down-fields: two triplets
of the scalar leptoquarks with the electric charges $Q^{em} = \pm 2/3$,
the $SU_C(3)$-octet of the neutral ``scalar gluons'' and four neutral
fields contained in $\phi^{(2)}_2$, $\phi^{(3)}_{15, 2}$, and
the $SU_L(2)$-up-partners of all these fields. All the scalar fields
are massive, some of them can be sufficiently heavy due to the mass scales
$\eta_1$ and $\eta_4$.

The Lagrangaian~(\ref{eq:LYukawa}) contains, in particular, the neutral
current interaction with the scalar fields. In the unitary gauge one
of the four neutral fields can be eliminated and the neutral current
interaction
of the fermions with the remained three scalar fields $\chi_1$, $\chi_2$
and $\omega_2$ can be written as

\begin{eqnarray}
{\cal L}^{scalar}_{N C} & = & {\cal L}_{N C} (\chi_1) +
{\cal L}_{N C} (\chi_2, \omega_2),  \nonumber \\
{\cal L}_{N C} (\chi_1) & = & - \frac{\chi_1}{\eta}
\big ( \bar Q_a M_{Q_a} Q_a + \bar \ell_a M_{\ell_a} \ell_a \big ),
\label{eq:LHi1} \\
{\cal L}_{N C} (\chi_2, \omega_2) & = &
\frac{\chi_2 - i \omega_2 ( \tau_3 )_{a a}}{\eta}
\frac{1}{2 \sin 2\beta} \nonumber \\
& \times & \big \lbrace
\bar Q^L_a \left [ (1 - 2 \cos 2 \beta) M_{Q_a} +
K^L_a M_{\ell_a} ( K^R_a )^+ \right ] Q^R_a \nonumber \\
& + & \bar \ell^L_a \left [ (-1 - 2 \cos 2 \beta) M_{\ell_a} +
3 (K^L_a)^+ M_{Q_a} K^R_a \right ] \ell^R_a
\big \rbrace + h.c. ,
\label{eq:LHi2}
\end{eqnarray}

\noindent where
$(M_{Q_a})_{p q} = m_{Q_{a p}} \delta_{p q}$ and
$(M_{\ell_a})_{p q} = m_{\ell_{a p}} \delta_{p q}$ are the
diagonal quark and lepton mass matrices, $\beta$ is the
$\phi^{(2)}$ - $\phi^{(3)}_{15}$ mixing angle,
$\tan \beta = \eta_3 / \eta_2$.

One can see from~(\ref{eq:LHi1}) that $\chi_1$ is the usual Higgs field
of the Standard Model. The interaction~(\ref{eq:LHi2}) contains the
flavour diagonal interactions represented by the first terms
and by the diagonal matrix elements of the second terms in the square
brackets. Besides, the interaction~(\ref{eq:LHi2}) contains, in general,
also the flavour changing neutral current (FCNC) interactions
described by the nondiagonal matrix elements of the second terms in
the square brackets. It is
interesting that FCNC interaction of the leptons depends on
the masses of the quarks and vice versa. It should be noted that
the ``dangerous'' FCNC interactions can be sufficiently supressed by the
smallness of the corresponding nondiagonal elements of the
$K^{L,R}_a$-matrices
and (or) by the large masses of the $\chi_2$ and $\omega_2$ fields.
It is pertient to note the particular case of
$A^{L,R}_{Q_a} = A^{L,R}_{\ell_a}$ when
$K^{L,R}_a = I$, $C_\ell = C_Q$ (without any contradiction with
the experiment) and the FCNC interactions in~(\ref{eq:LHi2}) are
absent at all.

The doublets $\phi^{(2)}$, $\phi^{(3)}_i$, $i = 1$, $2$, $\ldots$, $15$
interact with the photon, $W^\pm$- and $Z$-bosons and, in general,
will contribute into the radiative correction parameters $S$, $T$ and
$U$ of Ref.~\cite{Peskin}. The immediate calculation of the contributions
$S^{(\phi)}$, $T^{(\phi)}$ and $U^{(\phi)}$ of one scalar doublet
$\phi$ with the standard model hypercharge $Y^{SM}$ into $S$, $T$ and
$U$ parameters yields

\begin{eqnarray}
S^{(\phi)} & = & - \frac{Y^{SM}}{12 \pi} \, \ln \frac{m_1^2}{m_2^2} ,
\label{eq:Sphi} \\
T^{(\phi)} & = & \frac{1}{16 \pi s_W^2 c_W^2 m_Z^2} \,
\left [ m_1^2 + m_2^2 - \frac{2 m_1^2 m_2^2}{m_1^2 - m_2^2} \,
\ln \frac{m_1^2}{m_2^2} \right ],
\label{eq:Tphi} \\
U^{(\phi)} & = & \frac{1}{12 \pi} \,
\bigg [ - \frac{5 m_1^4 - 22 m_1^2 m_2^2 + 5 m_2^4}{3 (m_1^2 - m_2^2)^2}
\nonumber \\
& + & \frac{m_1^6 - 3 m_1^4 m_2^2 - 3 m_1^2 m_2^4 + m_2^6}{(m_1^2 - m_2^2)^3}
\,
\ln \frac{m_1^2}{m_2^2} \bigg ],
\label{eq:Uphi}
\end{eqnarray}

\noindent where $m_1$ and $m_2$ are the masses of the up and down
components of the doublet $\phi$. It should be noted that the
contribution~(\ref{eq:Sphi}) differs essentially from the standard
fermion doublet contribution~\cite{Peskin}. In particular, it does
not contain the mass independent term $(6 \pi)^{-1}$ and is not
positive defined. The contribution~(\ref{eq:Tphi}) coincides with
that from standard fermion doublet, whereas the contribution~(\ref{eq:Uphi})
is less than fermionic one by a factor two.

Applying the formulae~(\ref{eq:Sphi})--(\ref{eq:Uphi}) to the multiplets
$\phi^{(2)}$ and $\phi^{(3)}$ one should keep in mind that the multiplet
$\phi^{(3)}$ contains, in general, eight doublets with $Y^{SM} = 1$,
three doublets with $Y^{SM} = 7/3$, three doublets with $Y^{SM} = - 1/3$
and the doublet $\phi_{15}^{(3)}$ which together with the doublet
$\phi^{(2)}$ forms the standard Higgs doublet and an additional scalar
doublet with $Y^{SM} = 1$. The resulting contribution of $\phi^{(2)}$-
and $\phi^{(3)}$-multiplets into $S$, $T$ and $U$ defined
by these values of the hypercharge and by the masses of the corresponding
up and down fields will satisfy the current constarints~\cite{PDG} on
$S$, $T$, $U$ if the mass splittings of the scalar doublets are
sufficiently small. In particular case of the degenerate scalar doublets
their contributions into $S$, $T$, $U$, unlike the standard fermion doublet
case, are equal to zero.

The mass relation~(\ref{eq:mr}) gives the limit on the $Z'$-mass.
Using the experimental values of $G_F$, $m_W$ and $\alpha(m_Z)$~\cite{PDG}
we have $s_W^2 = 0.2298 \pm 0.0014$. Then
taking the most stringent limit $M_C \ge 10^5 \div 10^6 \, GeV$
resulting from $Br (K^0_L \rightarrow \mu e) < 0.94 \cdot 10^{-10}$~\cite{PDG}
into account and using the experimental values
$\alpha_s(m_Z) = 0.117 \pm 0.005$~\cite{PDG}
we evaluate $s_S^2$ and $\sigma$ from~(\ref{eq:sS}) and~(\ref{eq:s}) for
$M_C = 10^{6} \div 10^{14} \, GeV$ and $M' \ll M_C$ (see Table 1).
For these values of the $\sigma$ the relation $m_{Z'}/m_Z$
and $\sin \theta_m$ as functions of the $\Delta \rho_0 \equiv \rho_0 - 1$
are presented on Fig.1.
Taking the current value $\rho_0 = 1.0004 \pm 0.0022 \pm 0.002$~\cite{PDG}
into account we see from~Fig.1 that the
$Z'$-boson may be rather light.
Thus for $M_C = 10^6 \, GeV$
and $\Delta \rho < 0.002$
we get the limits  $m_{Z'} > 5 \, m_Z$
on $Z'$-mass and $\theta_m < 0.01$ on the $Z$-$Z'$ mixing
angle. This upper limit on $\theta_m$ is compatible with those obtained
in the extended gauge models~\cite{LL,R,A}.

Using the structure~(\ref{eq:Lnc})~-~(\ref{eq:JZ2}) of the neutral current
interaction we have calculated in the tree approximation the cross
sections $\sigma_{\bar f f} = \sigma (e^+ e^- \rightarrow \gamma, Z, Z'
\rightarrow \bar f f)$. The leptonic cross section $\sigma_{\bar \ell \ell}$
is found to be less than the one predicted by the SM.
This effect is due to
the destructive $\gamma$-$Z'$-interference~\cite{PSch}. The magnitude
of this deviation depends on the MQLS-model parameters $m_{Z'}$ and $M_C$.
For instance, at $M_C = 10^6 \, GeV$ the relative deviation
$\delta_{\bar \ell \ell} = ( \sigma_{\bar \ell \ell} -
\sigma_{\bar \ell \ell}^{SM}) / \sigma_{\bar \ell \ell}^{SM}$ of the
leptonic cross section $\sigma_{\bar \ell \ell}$ from the SM prediction
$\sigma_{\bar \ell \ell}^{SM}$ at the TRISTAN energies ($\sqrt s \simeq
60 \, GeV$) is about $\delta_{\bar \ell \ell} \simeq - 6 \%$
for $m_{Z'} = 4 m_Z$ and $\delta_{\bar \ell \ell} \simeq - 1 \%$
for $m_{Z'} = 10 m_Z$.
These deviations are of the same order as the experimental errors
of the leptonic cross section measurements at TRISTAN. The current
values of the measured leptonic cross sections at TRISTAN are still
slightly lower but they are consistent with the SM prediction \cite{S}.
The measurements of the leptonic cross sections with $2 \%$ accuracy
which is to be achieved soon at TRISTAN can give additional limits
on $m_{Z'}$ and $M_C$.
At the LEP200 energies ($\sqrt s \simeq
200 \, GeV$) these deviations are significantly larger and reach the values
$\delta_{\bar \ell \ell} \simeq - 60 \%$ and
$\delta_{\bar \ell \ell} \simeq - 10 \%$ for $m_{Z'} = 4 m_Z$ and
$10 m_Z$ respectively.
Hence the measurements of the leptonic cross sections
$\sigma_{\bar \ell \ell}$  at LEP200 will allow either to observe the
manifestation of the $Z'$-boson originated from the four-color
quark-lepton symmetry or to obtain the more stringent limits on
the MQLS-model parameters~$m_{Z'}$ and~$M_C$.

The hadronic cross section $\sigma_h = \sum_q \sigma_{\bar q q}$
is found to be somewhat more than that predicted by the SM
but this deviation is smaller than that in the leptonic case.
For instance, at $M_C = 10^6 \, GeV$ the relative deviation $\delta_h =
(\sigma_h - \sigma_h^{SM}) / \sigma_h^{SM}$ at TRISTAN energies
is only about $\delta_h \simeq 1.2 \%$
for $m_{Z'} = 4 m_Z$ and does not exceed the experimental errors
at TRISTAN. At the LEP200 energies
this deviation is about $\delta_h \simeq 20 \%$ for $m_{Z'} = 4 m_Z$
and $\delta_h \simeq 1.4 \%$ for $m_{Z'} = 10 m_Z$ and hence
at $5 \%$ accuracy may be also observable if $m_{Z'} \sim
400 \div 600 \, GeV$. As seen the measurements of the
leptonic cross sections are more favourable for the search
for the possible manifestations of the extra $Z'$-boson than
the measurements of the hadronic cross sections.

In conclusion we resume the results of the work. The minimal
quark-lepton symmetry model of the unification of the strong and
electroweak interactions as a minimal extension of the Standard
Model containing the four-color quark-lepton symmetry is
proposed. The new relation between the masses of the $W$-, $Z$ -
and $Z'$-bosons is obtained. The existence of a rather light extra
$Z'$-boson originated from the four-color quark-lepton symmetry
is shown to be compatible with the current electroweak data.

Taking the structure of the neutral current interaction specified
by the MQLS-model into account the cross sections of $e^+ e^-$
annihilation into leptons and hadrons are calculated and analysed.
The deviations of these cross sections from the SM predictions are
shown to be rather significant and available for the measurements
at LEP200 and future colliders. These measurements will allow
either to observe the manifestation of the $Z'$-boson originated
from the four-color quark-lepton symmetry or to obtain the more
stringent limits on the mass of $Z'$-boson and on the mass scale
of the four-color symmetry breaking.

\bigskip

{\bf Acknowledgments}

\bigskip

The author is grateful to N.V.~Mikheev, A.A.~Gvozdev and
L.A.~Vassilevskaya for discussions of the results and
to A.Ya.~Parkhomenko for the help in the work.
The work was supported by the Russian Foundation for Fundamental
Research (Grant No. 93-02-14414).

\newpage

\begin{table}[h]
\caption{The strong mixing angle $\sin^2 \theta_S$ and the parameter
$\sigma$ depending on the mass scale $M_C$ in the MQLS-model.}

\vspace{10mm}

\begin{center}
\begin{tabular}{ccc}\hline
$M_C$, $GeV$ & $\sin^2 \theta_S$ & $\sigma$   \\ \hline
$10^6$       &  0.130            &  0.216     \\
$10^{10}$    &  0.213            &  0.297     \\
$10^{14}$    &  0.297            &  0.380     \\ \hline
\end{tabular}
\end{center}

\vspace{80mm}

\end{table}

\newpage

{\Large\bf Figure caption}

\bigskip

\begin{quotation}
\noindent Fig. 1. Mass relation $m_{Z'}/m_Z$ and $\sin \theta_m$ as functions
         of the $\Delta \rho_0$ in MQLS-model: a)~$M_C=10^6 \, GeV$,
         b)~$M_C=10^{10} \, GeV$, c)~$M_C=10^{14} \, GeV$.
\end{quotation}








\newpage

\begin{thebibliography}{7}
\bibitem{PSm}
   J.C.~Pati and A.~Salam, Phys.~Rev. D10~(1974)~275.
\bibitem{KM}
   A.V.~Kuznetsov and N.V.~Mikheev, Phys.~Lett. B329~(1994)~295.
\bibitem{VW}
   G.~Valencia and S.~Willenbrock, preprint ILL-(TH)-94-17 (1994)
   (to appear in Phys.~Rev.~D).
\bibitem{Peskin}
   M.E.~Peskin and T.~Takeuchi, Phys.~Rev. D46~(1992)~381.
\bibitem{PDG}
   Particle Data Group, L.~Montanet et~al., Phys.~Rev. D50~(1994)~1173.
\bibitem{LL}
   P.~Langacker and M.~Luo, Phys.~Rev. D45~(1992)~278.
\bibitem{R}
   S.~Riemann, DESY preprint DESY 92-143 (1992).
\bibitem{A}
   G.~Altarelli et al., CERN preprint CERN-TH.6947/93 (1993).
\bibitem{PSch}
   A.A.~Pankov and J.S.~Satsunkevich, Yad.~Fiz. 47~(1987)~1333.
\bibitem{S}
   M.~Sakuda, KEK preprint 93-124 (1993).
\end{thebibliography}
\end{document}